\documentstyle[prl,aps,epsfig]{revtex}
\begin{document}
%\draft
%\tightenlines
\title{Quantum dots in magnetic fields:
       thermal response of broken symmetry phases}

\author{D.J. Dean$^{1,2}$, M.R. Strayer$^{1,2}$, 
and J.C. Wells$^{1,3}$}

\address{
$^1$Center for Engineering Science Advanced Research \\ 
Oak Ridge National Laboratory, Oak Ridge, Tennessee 37831 USA\\
$^2$Physics Division, 
Oak Ridge National Laboratory, Oak Ridge, Tennessee 37831 USA\\
$^3$Computer Science and Mathematics Division,
Oak Ridge National Laboratory, Oak Ridge Tennessee 37831 USA
}

\maketitle

\begin{abstract}
We investigate the thermal properties of 
circular
semiconductor
quantum dots in high magnetic fields 
using finite temperature Hartree-Fock
techniques. We demonstrate that for a given magnetic
field strength quantum dots undergo various shape 
phase transitions as a function of temperature,
and we outline possible observable consequences. 
\end{abstract}


The quantum mechanical characteristics (i.e. ground- and
excited-state energy levels) of semiconductor quantum dots
were recently investigated as a function of increasing
magnetic field and electron number \cite{Stewart97,Kouwenhoven97}.
Unique structures that break rotational symmetry within a quantum 
dot and develop as a function of increasing magnetic field were 
also reported in theoretical investigations \cite{Reimann99}
and associated with experimental observations of
charge-density redistribution \cite{Reimann99,Oosterkamp99}. 
In this paper, we explore the thermal characteristics of 
the broken-symmetry phase of the charge-density distribution,
showing that temperature can be used to induce transitions
between phases of differing spatial symmetry.

Thermal properties of, and transitions in, 
strongly correlated quantum mechanical systems
have been studied in various contexts. 
Perhaps the most obvious are the normal-conducting to 
super-conducting transitions in condensed matter systems.  
Many nuclear systems exhibit an intrinsic deformation in their
ground states, and deformed-to-spherical 
transitions have been studied in many
models (see for example Refs.~\cite{izc98,dean95,white00}).
As a result, observable consequences, such as enhanced moments 
of inertia, decreasing quadrupole transition moments, and 
decreasing correlated pair-transfer amplitudes may be investigated. 

In studies of many-body phenomena in quantum dots,
experimental efforts have focused on mapping the magnetic field
dependence of their
ground-state
structure by measuring the chemical potential 
via capacitance spectroscopy \cite{Ashoori}.
Cusps and steps in the chemical potential were found to clearly
separate different ranges of magnetic fields \cite{Ashoori,Oosterkamp99}.
These features were identified with phase transitions in the charge
density of the quantum dot. At magnetic field strengths on
the order of a few tesla, all electrons become
spin-polarized initiating the maximum density droplet (MDD)
phase \cite{Ashoori}, in which
the density is constant and homogeneous at the
maximum value that can be reached in the lowest Landau level.
The stability of the MDD is determined by a competition among the
kinetic energy, external confinement, the Coulomb repulsion between
electrons, and the attraction created by the Coulomb exchange term.
For increasing magnetic field, the charge-density distribution
of the droplet reconstructs \cite{Chamon} with a ring of electrons
breaking off from the MDD phase. 
This edge reconstruction has been shown via mean-field \cite{muller}
and density functional theory \cite{Reimann99} calculations
to result from a rotational symmetry-breaking phase transition
from the MDD to a Wigner molecule or Wigner crystal phase. 
These calculations are in good qualitative agreement with recent
experimental results where instabilities of the MDD state
and other transitions in the high magnetic field region
were accompanied by a redistribution of the
charge density \cite{Oosterkamp99}.

Most of the experimental investigation of many-body phenomena
in quantum dots 
was performed at very low temperatures on the 
order of 100 mK and focused on the evolution of the ground state
of the quantum dot as a function of increasing magnetic field 
and electron number \cite{foxman94,klein96}.
The thermal characteristics of these systems are largely
unexplored, especially for large values of the magnetic field. 
We explore the thermal response of the broken-symmetry phase
at constant magnetic field and electron number and demonstrate
that transitions among a number of distinct,
broken-symmetry phases occur as the temperature is increased.

In describing the ground-state and low-lying (intra-band) excitations of
the $N$-electron semiconductor nanostructures, it is often sufficient to
restrict consideration to the conduction band using the effective-mass
approximation \cite{Jacak}.
We consider the problem of N electrons of effective mass $m^*$
in a plane, $(x,y)$, confined by an external parabolic potential,
$V(r) = \frac{1}{2} m^* \omega_0^2 r^2 $, and subject to a
strong magnetic field $\vec{B} = B_0 \vec{e}_z $.
We consider the Zeeman splitting but neglect the
spin-orbit interaction.  The Hamiltonian for such a system is
\begin{eqnarray}
\hat{H} &=&
 \sum_i \left[
   \frac{\left( \vec{p}-\frac{e}{c}\vec{A} \right)^2}{2m^*}
 + V(r_i)
 + \frac{g^* \mu_B}{\hbar} \vec{B} \cdot \vec{S}_i  \right] \nonumber \\
 &+& \sum_{i<j} \frac{e^2}{\varepsilon |\vec{r}_i - \vec{r}_j | }\;,
\label{intraband}
\end{eqnarray}
where the vector potential is $\vec{A}(\vec{r}_i) = (B_0/2)(-y_i,x_i,0)$,
$g^*=0.54$, $\varepsilon=12.9$, $m^*=0.067 m_e$, and
$\hbar \omega_0 = 3$~meV. 

We solve this equation at the finite-temperature Hartree--Fock level. 
This is the initial step to account for electron-electron
correlations within the quantum dot and suffices for a qualitative
discussion of phase transitions and thermal behavior. Hartree-Fock
is implemented across nuclear and atomic physics as a first step 
towards the solution of the quantum many-body problem \cite{sdn98}.
Since the Hartree-Fock equations are nonlinear, the self-consistent
potential obtained from their solution will not necessarily show the
same symmetries as the original Hamiltonian. Of course, the exact
wave will maintain the symmetries of the original Hamiltonian. 
As a point of reference, an estimate of the relative uncertainty
in total angular momentum at high magnetic field strengths 
was shown to be roughly $10\%$ by the authors of Ref.~\cite{muller}.

The equations describing the static mean field (i.e., Hartree-Fock)
at finite temperature for an $N$-electron system are
\begin{equation}
[\hat{K} + \hat{W}(\beta) - E_\alpha] | \phi_\alpha(\beta) = 0
\end{equation}
where $E_\alpha$ are the single particle energies associated with
the single-particle wave functions $\phi_\alpha$, and
$\hat{K}$ is the one-body kinetic-energy operator. 
The one-body Hartree--Fock field, $\hat{W}(\beta)$, is obtained
in terms of the Hamiltonian (\ref{intraband}) and the one-body density
matrix, $\hat{\rho}(\beta)$ as,
\begin{equation}
\hat{W}(\beta) = {\rm Tr}[\hat{\rho}(\beta)\hat{H}]\; .
\label{eq2}
\end{equation}
The single-particle wave functions are denoted as
$|\phi_\alpha(\beta)\rangle$,
and  $\beta = 1/kT$ is the inverse temperature.
The representation of the one-body density is given in terms of the
occupation numbers $n_\alpha(\beta)$,
$ \rho(\beta) = \sum_{\alpha}
n_\alpha(\beta)\mid\phi_\alpha(\beta)\rangle\langle\phi_\alpha(\beta)\mid$,
and the Fermi--Dirac occupation numbers at chemical potential $\mu$ are
$n_\alpha(\beta) = 1/\left[e^{\beta(E_\alpha - \mu)} + 1\right]\;$.
Note that Tamura and Ueda \cite{tamura97} studied the
number fluctuations $\langle \left(\delta N\right)^2\rangle$ 
of a quantum dot as a function of $N$ for different
field strengths and found that the fluctuations
were rather small (on the order of 0.1) in finite-temperature
Hartree--Fock applications. We expect qualitative
agreement with experiment as Hartree--Fock is known to 
overestimate the magnetic field strength at which
transitions occur \cite{klein95,klein96,Reimann99}. 
However, for a fixed field strength, experimental
evidence of the thermal transitions 
should still be qualitatively visible. 

We use a Fock--Darwin basis expansion to solve the finite-temperature
Hartree-Fock equations. 
Since we use a high ($\approx 12$~T) magnetic field, we consider only angular
momentum states with the $n=0$ principal quantum number. The
electrons carry spin, and so our states are labeled by 
$k=\{l_k,s_k\}$, where $l_k$ is the angular momentum projection
of the $k$-th state and $s_k$ is the spin of that state. We found
convergence using fifty states for the $N\le8$ systems. 
We also checked our zero-temperature results with other
publications \cite{muller} for various numbers of 
electrons in the dot and found satisfactory agreement. 

\begin{figure}
\begin{center}
\vskip0.26in
\epsfig{file=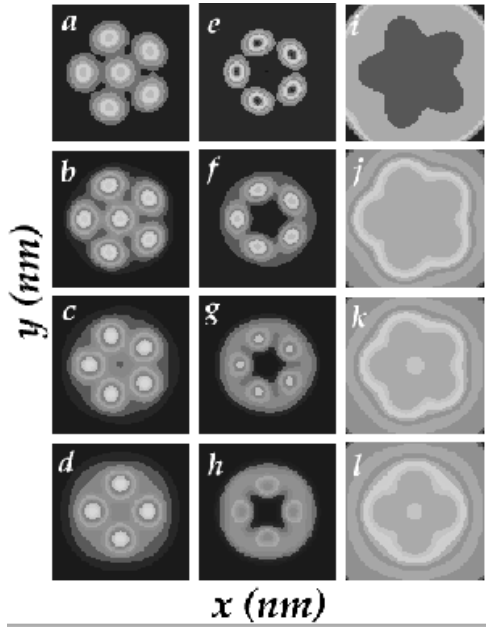,width=2.5in}
\end{center}
\caption{Panels a-d, the charge density; panels e-h, the
angular momentum density (in the $z$-direction); panels
i-l, the spin density.  In each panel, $-8$~nm$\le x,y \le 8$~nm. }
\label{figure_density}
\end{figure}

We begin the discussion of our results by investigating the electron
charge, angular momentum, and spin densities
as a function of increasing temperature for the $N=6$ system
at $B_0=12.15$~T.  We show densities at representative 
temperatures of 3.87~K (the low temperature limit, $\nu=6$), 
11.97~K (before the first phase transition, $\nu=6$), 
13.65~K (in the second phase, $\nu=5$), and 14.32~K (in the third
phase, $\nu=4$), where $\nu$ is the number of definable high-density 
regions (or vortices) in the charge density plots. 
The density, shown in Fig.~\ref{figure_density}a-d,
begins as a fairly well-defined Wigner crystal at 3.87~K, which
exhibits some degree of thermal broadening at 11.97~K. The $\nu=5$
and $\nu=4$ phases continue to show a similar amount of density
in the remaining vortices, while the density of the 
thermally dissipated vortices have effectively spread through the entire 
dot. The angular momentum along the $B$-field direction, 
shown in Fig.~\ref{figure_density}e-h,
exhibits well-defined structures at low temperatures which tend 
to decrease rapidly as one moves through the various phases. 
Although the high charge density regions in the $\nu=4$ phase 
are still well defined, the angular momentum in this high-temperature
phase has nearly washed out. Finally, we show in 
Fig.~\ref{figure_density}i-l the spin density defined as
$\rho_s(x,y)=\left[\rho_{\uparrow}(x,y)-\rho_{\downarrow}(x,y)\right]/
\left[\rho_{\uparrow}(x,y)+\rho_{\downarrow}(x,y)\right]
$, where $\rho_{\uparrow}$ ($\rho_{\downarrow}$) refer to the
spin up (down) density. At these temperatures, little appreciable
spin 
depolarization
occurs and the spin density remains above 0.8
for the entire region where there is appreciable charge density.

\begin{figure}
\begin{center}
\epsfig{file=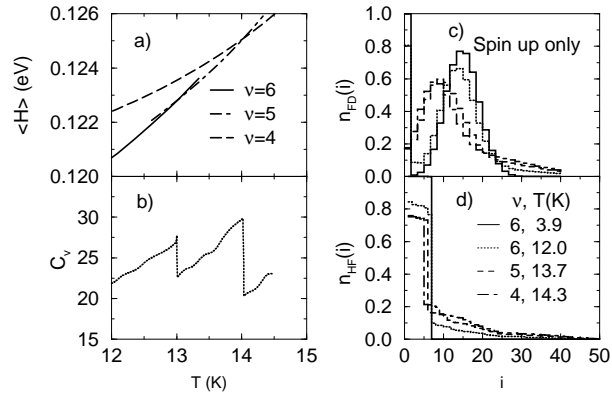,width=3.2in}
\end{center}
\caption{
a) Expectation value of the energy as a function of temperature showing
the three phases as discussed in the text. b) The specific heat
for the lowest-energy configuration of the quantum dot as a 
function of temperature. c) Occupation of the Fock--Darwin states.
d) Occupation of the Hartree--Fock states. }
\label{figure_ene}
\end{figure}

The 
suddenness of the
phase transitions seen in Fig.~\ref{figure_density} become
quite evident when the internal energy of the quantum dot 
is plotted as a function of the temperature.  We show
the three phases of the dot in Fig.~\ref{figure_ene}a. Note
that the $\nu=6$ phase exists as an excited configuration when
the most probable Hartree--Fock solution is the $\nu=5$ phase. 
Similarly, the $\nu=4$ configuration exists as a possible 
excited configuration of the system even at fairly low temperatures. 
The specific heat, $C_v=d\langle H\rangle/dT$ (with T in units of eV),
is shown in Fig.~\ref{figure_ene}b. Clearly, when the energy
undergoes a phase transition, the specific heat shows a sharp
structure. This occurs since the energy is piece-wise continuous
along the three phases. 

These calculations suggest that the quantum dot exhibits a band structure
of many-body levels. The low-temperature states all have the 
same intrinsic shape characteristic (the same vortex structure). 
As we increase the temperature, other $\nu$ phases become accessible. 
At the point when two bands of different intrinsic character 
cross in energy, we find a phase transition. Similar phenomena
are found in nuclear physics, where at higher nuclear excitation
energies the eigenstates of the system may be of a different intrinsic 
deformation when compared to states of the
ground-state band \cite{baktash98}.

We also find generally that the 
exchange energy decreases as a function of increasing 
temperature but with differing slopes in the 
different phases. In the region of 
phase transitions, the ratio of the exchange energy to 
the direct energy decreases from 0.38 
(at 11.5 K)to 0.32 (at 13 K) in the $\nu=6$ phase. 
In the $\nu=5$ phase, this ratio decreases
from 0.32 at $T=13$~K to 0.28 at $T=13.8$~K, and from
0.28 at $T=13.8$~K to 0.26 at $T=14.5$~K in the $\nu=4$ phase.
The slopes of the decreasing ratio are different across
the three phases, with $\nu=6,5$ being the largest and 
$\nu=4$ more gradual.

The occupation probabilities of the Fock--Darwin states, $n_{FD}$,
for the temperature conditions of Fig.~\ref{figure_density}
are shown in Fig.~\ref{figure_ene}c. In the low-temperature phase
(3.9 K, solid line), the $l=0$ state is occupied, while
an edge reconstruction has occurred for the remaining 
five electrons.  As we increase the temperature to 
12 K (dotted line), we see a spreading of the
occupations in both the low and high angular momentum
channels with about 0.8 particles in the $l=0$ state. The
$\nu=5$ phase brings a dramatic decrease of occupation in the
$l=0$ state and a shift to lower angular momentum for the 
reconstructed edge. This trend continues after the $\nu=4$ transition.

Signatures of the 
density transitions that we have seen also appear
in the Hartree--Fock occupations $n_{HF}$ as shown in Fig.~\ref{figure_ene}d.
At low temperatures (3.9 K) the familiar step-function behavior
is evident from the figure. At 12 K in the $\nu=6$ phase, we
see a decrease of occupation to roughly 0.8 in the lowest 
six Hartree--Fock levels and a spreading to higher energy states. 
As the system undergoes the transition to $\nu=5$, we
see only five Hartree--Fock levels significantly filled 
(with $n_{HF}>0.7$), and finally in the
$\nu=4$ phase, only four Hartree-Fock levels are significantly 
filled.  The occupation number-spreading, which is due to thermal
excitation of the system, is enhanced significantly when 
the system undergoes a phase transition. 

\begin{figure}
\begin{center}
\epsfig{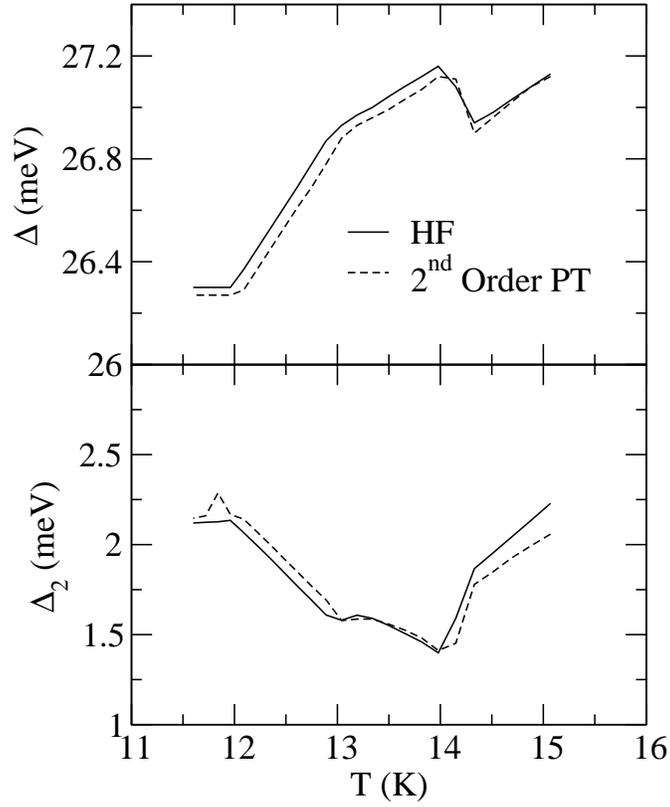}
\end{center}
\caption{a) Chemical potential, $\Delta$ and b) inverse compressibility,
$\Delta_2$ for the $N=6$ system as a function of the temperature. The
Hartree-Fock results are shown as a solid line, and the second-order 
perturbation theory results are given by the dashed line.}
\label{xmu_fig}
\end{figure}

The phase transitions that we have shown in the preceding discussion
have definite observable consequences. In Fig.~\ref{xmu_fig}a we 
plot the chemical potential, that is, the 
separation energy $\Delta(N,T)=E_{N}(T)-E_{N-1}(T)$ to remove
a particle from the quantum dot at a given temperature. Since
this is an energy difference, $\Delta(N,T)$ will be influenced
by transitions within both the $N=6$ and $N=5$ systems, and
we expect changes in slope at the transition points. 
The $\nu=5$ to $\nu=4$ transition in the $N=5$ dot occurs
at roughly 12~K, causing a sharp rise in $\Delta(N=6)$. 
A slope change in $\Delta(N=6)$ is seen at $\approx 13$~K, where
the $\nu=6$ to $\nu=5$ transition occurs in the $N=6$ system. 
The decrease from $14-14.3$~K occurs when the $N=6$ system
makes the transition from $\nu=5$ to $\nu=4$. A final change in
slope occurs when the $N=5$ dot makes the transition from
$\nu=4$ to $\nu=3$. For the brief temperature interval when
the $N=6$ and $N=5$ dots are in the same $\nu$ phase, we see
a decrease in the chemical potential. 

We observe similar changes in the inverse compressibility,
$\Delta_2(N,T)=E_{N+1}(T)-2E_N(T)+E_{N-1}(T)$. 
This quantity has been measured for quantum dots in 
low magnetic fields \cite{d2} and studied 
in Hartree--Fock theory for ground-state properties \cite{levit99}.
In our case,
the $N=7,6$, and $5$ dots participate in the observable. We
again notice strong effects as one passes through transition
points in either of the three systems contributing to the
observable.  Figure~\ref{xmu_fig}b shows $\Delta_2(N=6)$ as
a function of temperature. Before transitions occur, $\Delta_2$
remains fairly constant. A large decrease begins at 
12 K, where the $N=5$ system undergoes its first transition. 
Interestingly, $\Delta_2$ increases significantly 
when the $N=6$ and $N=5$ dots are in the $\nu=4$ phase. 

In order to investigate the sensitivity of our results to the 
approximations made (namely the Hartree-Fock approximation) 
we extended our studies to include the second-order
perturbative correction to the Hartree-Fock energy. This correction
to the energy is given by 
\begin{equation}
\Delta E_2 = \frac{1}{4}\sum_{ab\le \epsilon_F} \sum_{rs > \epsilon_F}
\frac{\mid \langle a b \mid \bar{v} \mid r s \rangle \mid^2}
{\varepsilon_a + \varepsilon_b -\varepsilon_r -\varepsilon_s}
\end{equation}
where we restrict the sums to be below and above the Fermi energy surface,
$\epsilon_F$. In this expression, the Hartree-Fock states are given by
$a,b,r,s$, with associated single-particle energies
$\varepsilon_a, \varepsilon_b, \varepsilon_r, \varepsilon_s$, and 
$\langle ab\mid\bar{v}\mid rs\rangle$ are the antisymmetrized two-body
Hartree-Fock matrix elements of the original two-body interaction of 
the Hamiltonian.
We note that the absolute value of $\Delta E_2$ in our calculations is
always less than $1\%$ of the total energy, and in fact only slightly
changes the observable quantities, as is evident from Fig.~\ref{xmu_fig},
where the second-order perturbative results are also shown for $\Delta$
and $\Delta_2$. 

\begin{figure}
\begin{center}
\epsfig{file=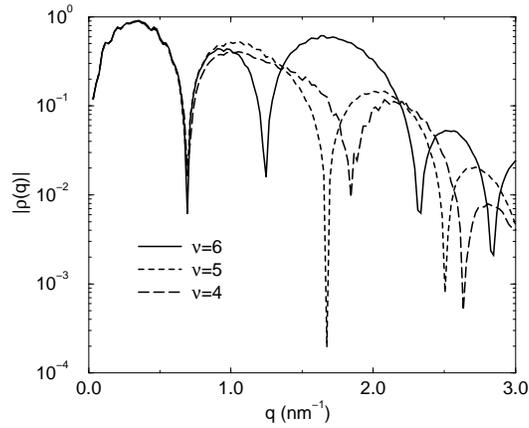,width=3.0in}
\end{center}
\caption{
The form factor in the 
three phases, $\nu=6$ (11.9 K), $\nu=5$ (13.6 K), and $\nu=4$ (14.3 K). }
\label{figure_four}
\end{figure}

In addition to measurements of energy differences, 
one should be able to experimentally probe the thermal 
phase transitions using far-infrared spectroscopy
and X-ray scattering. 
Far-infrared spectroscopy was used to investigate the
excitations of InAs quantum dots as a function of the electron
number per dot \cite{fricke96}.
Grazing incidence 
X-ray scattering was recently used to generate
a full structural characterization of quantum dots\cite{metzger98},
including information on the elastic form factor. 
While these experiments were carried out at very low temperatures,
it is conceivable that one could study the thermal response
of quantum dots using X-ray or far-infrared scattering. A Fourier
transform of the charge density produced in our calculations
gives the elastic form factor that
can be used to characterize the dot. 
We show in Fig.~\ref{figure_four} 
the form factor, $\mid\rho(q)\mid$, as 
a function of the momentum transfer vector 
$\mid q\mid =\sqrt{q_x^2+q_y^2}$.
A 
well-defined minimum is apparent 
at approximately $\mid q\mid=0.7$ev$^{-1}$
in all three phases.
This first minimum is related to the size of the
dot and clearly does not change in the three phases. This is apparent
also from a close inspection of Fig.~\ref{figure_density}: while the
internal structure changes significantly as a function of 
increasing temperature, the dot size does not change. The
height of the second maximum slightly increases as in phase 
$\nu=5$ before decreasing in the $\nu=4$ phase. The position of the
second minimum increases as a function of $q$ significantly and
would be a distinguishing feature in X-ray or far-infrared scattering
experiments used to probe the thermal phases of a quantum dot. 

We see in the above calculations that quantum dots 
exposed to high magnetic fields
undergo various
phase transitions as the temperature increases. 
Indeed, we have followed the $N=6$ system through the $\nu=3$
and $2$ phases as well. 
We have also studied these transitions at various 
magnetic-field strengths, and with differing numbers of 
electrons in the dot up to $N=20$,
with the results presented here being characteristic
of the general behavior.
Our purpose in this paper has been to demonstrate that thermal 
excitations of a quantum dot induce phase transitions  
within the dot that should have observable consequences.  
Thus, we have shown that temperature may also be used as a 
control to tune the characteristics of quantum dots. 
Inclusion of the full many-body correlations 
may slightly dampen some of these effects \cite{fbl01}, but signatures of the 
phase transitions should still appear in scattering experiments
and measurements of the inverse compressibility and
chemical potential of the quantum dot.

This research was sponsored by the Laboratory Directed Research 
and Development Program of Oak Ridge National Laboratory, under Contract 
No. DE-AC05-00OR22725 managed by UT-Battelle, LLC.  by the 
U.S. Department of Energy.

\bibliographystyle{try}

\begin{thebibliography}{99}

\bibitem{Stewart97}
D.R. Stewart, D. Sprinzak, C.M. Marcus, 
C.I. Duruoz, and J.S. Harris, Science {\bf 278}, 1784 (1997).

\bibitem{Kouwenhoven97}
L.P. Kouwenhoven, T.H. Oosterkamp, M.W.S. Danoesatro,
M. Eto, D.G. Austing, T. Honda, and S. Tarucha, 
Science {\bf 278}, 1788 (1997).


\bibitem{Reimann99}
S.M. Reimann, M. Koskinen, M. Manninen, and B.R. Mottelson,
Phys. Rev. Lett. {\bf 83}, 3270 (1999).

\bibitem{Oosterkamp99}
T.H. Oosterkamp, J.W. Janssen, 
L.P. Kouwenhoven, D.G. Austing, 
T. Honda, and S. Tarucha, Phys. Rev. Lett. {\bf 82}, 2931 (1999).

\bibitem{izc98}
F. Iachello, N.V. Zamfir, and R.F. Casten, 
Phys. Rev. Lett. {\bf 81}, 1191 (1998). 

\bibitem{dean95}
D.J. Dean, S.E. Koonin, K. Lanbanke, P.B. Radha, and 
Y. Alhassid, Phys. Rev. Lett. {\bf 74}, 2909 (1995).

\bibitem{white00}
J.A. White, S.E. Koonin, and D.J. Dean, Phys. Rev. {\bf C61},
34303 (2000).

\bibitem{Ashoori} 
R. C. Ashoori, Nature {\bf 379}, 413 (1996).

\bibitem{Chamon} 
C. de C. Chamon and X.G. Wen, 
Phys. Rev. B {\bf 49}, 8227 (1994).

\bibitem{muller} H.-M. M\"uller and S.E. Koonin,
Phys. Rev. B {\bf 54}, 14532 (1996).

\bibitem{foxman94}
E.B. Foxman, U. Meirav, P.L. McEuen, M.A. Kastner, 
O. Klein, P.A. Belk, D.M. Abusch, and
S.J. Wind, Phys. Rev. {\bf B50}, 14193 (1994).

\bibitem{klein96}
O. Klein, D. Goldhaber-Gordon, C. de C. Chamon, and M.A Kastner, 
Phys. Rev. {\bf B53}, R4221 (1996).

\bibitem{Jacak} 
L. Jacak, P. Hawrylak, and A. W\'ojs, {\it Quantum Dots}, 
(Springer, Berlin, 1997).

\bibitem{sdn98}
W. Satula, J. Dobaczewski, W. Nazarewicz, Phys. Rev. Lett. 
{\bf 81}, 3599 (1998)

\bibitem{tamura97}
H. Tamura and M. Ueda, Phys. Rev. Lett. {\bf 79}, 1345 (1997).

\bibitem{klein95}
O. Klein, C.D. Chamon, D. Tang, M. Abusch, D.M. Magder, U. Meirav,
X.G. Wen, M.A. Kastner, and S.J. Wind, Phys. Rev. Lett. {\bf 74}, 785 (1995).

\bibitem{baktash98}
D.G. Sarantites, D.R. LaFosse, M. Delvin, F. Lerma, V.Q. Wood, J.X. Saladin,
D.F. Winchell, C. Baktash, C.H. Yu, P. Fallon, I.Y. Lee, 
A.O. Macchiavelli, R.W. MacLeod, A.V. Afanasjev, and 
I. Ragnarsson, Phys. Rev. {\bf C57}, R1 (1998). 

\bibitem{d2}
U. Sivan, R. Berkovits, Y. Aloni, O. Prus, A. Auerbach, 
and G. BenYoseph, Phys. Rev. Lett. {\bf 77}, 1123 (1996);
S.R. Patel, S.M. Cronenwett, D.R. Stewart, A.G. Huibers, C.M. Marcus,
, C.I. Duruoz, J.S. Harris, K. Campman, and A.C. Gossard, 
Phys. Rev. Lett. {\bf 80}, 4522 (1998).

\bibitem{levit99}
S. Levit, D. Orgad, Phys. Rev. {\bf B60}, 5549 (1999).

\bibitem{fricke96}
M. Fricke, A. Lorke, J.P. Kotthaus, G. Medeiros-Ribeiro, 
and P.M. Petroff, Europhys. Lett. {\bf 36}, 197 (1996).

\bibitem{metzger98}
T.H. Metzger, I. Kegel, R. Paniago, A. Lorke, J. Peisl, J. Schulze, 
I. Eisele, P. Schittenhelm, and G. Abstreiter, Thin Solid Films {\bf 336},
1 (1998).

\bibitem{fbl01}
A.V. Filinov, M. Bonitz, and Yu. E. Lozovik, 
Phys. Rev. Lett. {\bf 86}, 3851 (2001).


\end{thebibliography}

\end{document}